# Enhancing Magnetic Coupling in MN4-Graphene via Strain Engineering

Mahnaz Rezaei[1], Jahanfar Abouie[2,*], Fariba Nazari[1,3,*]

[1]*Department of Chemistry, Institute for Advanced Studies in Basic Sciences,*

*Zanjan 45137-66731, Iran*

[2]*Department of Physics, Institute for Advanced Studies in Basic Sciences,*

*Zanjan 45137-66731, Iran*

[3]*Center of Climate Change and Global Warming, Institute for Advanced Studies in Basic*

*Sciences, Zanjan 45137-66731, Iran*

*Corresponding authors: jahan@iasbs.ac.ir, nazari@iasbs.ac.ir

**Abstract**

MN$_4$-embedded graphene (MN$_4$-G) layers, incorporating transition metal elements (M), represent a class of experimentally accessible two-dimensional materials with significant potential for stable nanoscale magnetization. In these systems, magnetic exchange interactions are primarily governed by Ruderman-Kittel-Kasuya-Yosida (RKKY) coupling, exhibiting an anomalously prolonged decay of $r^{-n}$, where $r$ is the M-M separation distance and $0.5 \leq n \leq 2$. This study investigates the impact of strain on the electronic and magnetic properties of MN$_4$-G layers using ab-initio density functional theory (DFT). A novel strain-engineering approach is developed by applying controlled tension or compression to the layers. Our findings reveal that strain significantly modulates the strength, amplitude, and decay rate of the RKKY coupling. Notably, the CoN$_4$-G layer demonstrates a pronounced enhancement in RKKY coupling strength, oscillation amplitude, and reduced decay rate under strain. Conversely, the CuN$_4$-G layer exhibits distinct behavior, maintaining decoupled spin chains and invariant electronic and magnetic properties despite applied strain. This work underscores the tunability of magnetic interactions in MN$_4$-G layers via strain engineering, providing insights into the design of strain-controlled magnetic materials for next-generation spintronic applications.

# I. Introduction

Quantum magnetism is an expansive field within condensed matter physics, captivating researchers across multiple scientific disciplines through its theoretical, experimental, and industrial applications. The quantum magnetism primarily originates from interacting magnetic moments of electrons, particularly in systems where unpaired electrons contribute significantly. In ferromagnets, the magnetic moments of neighboring atoms align parallel to each other, leading to a net magnetization, while in antiferromagnets, they align antiparallel resulting in a net staggered magnetization [1, 2]. These spin alignments are governed by exchange couplings, a purely quantum mechanical phenomenon resulting from the interplay between electrostatic repulsion among electrons and the Pauli exclusion principle [3]. Exchange couplings, which operate at atomic scales within a few angstroms, are critical in determining the magnetic properties of materials. They dictate how spins couple, leading to the emergence of magnetic order. The strength and nature of exchange couplings can vary, influencing whether a material exhibits ferromagnetism, antiferromagnetism, or other forms of magnetic ordering [4]. Understanding the mechanism behind exchange couplings is essential for tailoring materials with specific magnetic properties, which are crucial in numerous technological applications, ranging from data storage to quantum computing [5-10].

To uncover the origins of the macroscopic properties of strongly interacting materials, it is essential to examine how primary exchange couplings are influenced by external parameters. These parameters, including doping [11-14], intercalation [15-17], pressure [18-20], defect [21-25], electric field [26-29], and strain engineering [30-39], can be employed to modify material properties [40].

Among these methods, strain engineering stands out as a straightforward yet effective approach to tune the characteristics of two-dimensional (2D) materials [30-32]. This technique significantly influences their optoelectronic [33-36], electrocatalytic [37-39], and magnetic [41-43] properties by directly affecting lattice structures, which in turn alters the electronic structures. As a result, a wide range of physical properties can undergo substantial changes.

Extensive experimental and theoretical studies have explored how strain influences the magnetic properties of 2D materials, with a particular emphasis on the strength of magentic exchange coupling. For instance, researchers have successfully altered the strain in a Fe/CoO/MnO/MgO(001) layered structure by adjusting the thickness of the MnO layer, thereby achieving precise control over the exchange coupling within the bilayer [44]. Similarly, the introduction of artificial tensile strain, theoretically induced by substituting Ce in SmCo, enhances ferromagnetic exchange coupling [45]. In another theoretical study, it was revealed that tuning the in-plane lattice constant of CrSBr significantly modifies the interlayer magnetic exchange coupling which changes sign at a critical strain [46]. It has also been demonstrated that in IrMn/CoFeB bilayers, the Neel temperature, the magnetic moment of Mn atoms, and the magnetocrystalline anisotropy can be linearly controlled by strain [47].

Furthermore, it has been reported that variations in the Cr–I bond distance and Cr–I–Cr bond angle in the $CrI_3$ layer due to strain introduce a competition between direct exchange coupling and indirect superexchange coupling, resulting in a ground state with either a ferromagnetic or a Neel order [48].

Recent studies on $MN_4$-embedded graphene ($MN_4$-G) layers, where M represents transition metal elements, indicate that these experimentally accessible materials are promising candidates for achieving stable nanoscale magnetization [49-51]. Recently, we have demonstrated that the $MN_4$-

G layers with M = Mn, Fe, and Co, are 2D anisotropic magnetic layers where the exchange couplings between magnetic atoms are the Ruderman-Kittel-Kasuya-Yosida (RKKY) coupling [52]. In these layers the RKKY coupling exhibits an unusual prolonged decay of $r^{-n}$, where $r$ is the M-M separation distance, and $0.5 \leq n \leq 2$. We have also focused on designing model systems using bulk bcc-Fe doped with nonmagnetic atoms (H, B, C, N, O, and F) to explore the effects of various ligands on Fe-Fe exchange coupling [53]. Our results revealed that the local chemical environment surrounding the Fe atom plays a crucial role in influencing this coupling. We discovered that by fine-tuning the Fe-ligand bond lengths, we could effectively adjust the exchange coupling while maintaining the distance between magnetic atoms. Remarkably, even minor alterations in these bond lengths can lead to substantial and nonlinear shifts in exchange coupling values, underscoring the sensitivity of this relationship to slight changes in bond lengths [53].

Based on our findings, we anticipate that strain will alter the chemical environment surrounding M atoms in real systems ($MN_4$-G layers). While our previous study has examined the magnetic properties of $MN_4$-G layers, the effect of strain on the exchange coupling of M atoms remains largely unexplored. In this paper, we investigate how strain modifies the electronic structures and magnetic properties of the $MN_4$-G layers, where M = Mn, Fe, Co, Cu (see Fig. 1). We employ a specific method to apply strain by positioning atoms from one layer within the equilibrium structure of another layer, thereby inducing strain in the form of either tension or compression. Utilizing spin-polarized density functional theory (SP-DFT), we demonstrate that the strength, amplitude, and the decay rate of the RKKY coupling within most of these layers are significantly influenced by induced strain. For instance, we show that applying a certain amount of strain to the $CoN_4$-G layer significantly increases the strength and the amplitude of the RKKY oscillations while reducing the decay rate of the RKKY coupling. This reduction in the decay rate enhances the potential of these layers for spintronics applications.

This paper is organized as follows: In Section II, we introduce our approach to inducing strain in the MN$_4$-G layers. Section III examines the effects of strain on the electronic properties of MN$_4$-G layers. In Section IV, we present our findings on the magnetic properties of the strained MN$_4$-G layers, focusing on the effects of strain on the magnetic moment of the M atoms and the exchange coupling between them. Section V details our computational methods, and finally, Section VI summarizes our work and presents concluding remarks.

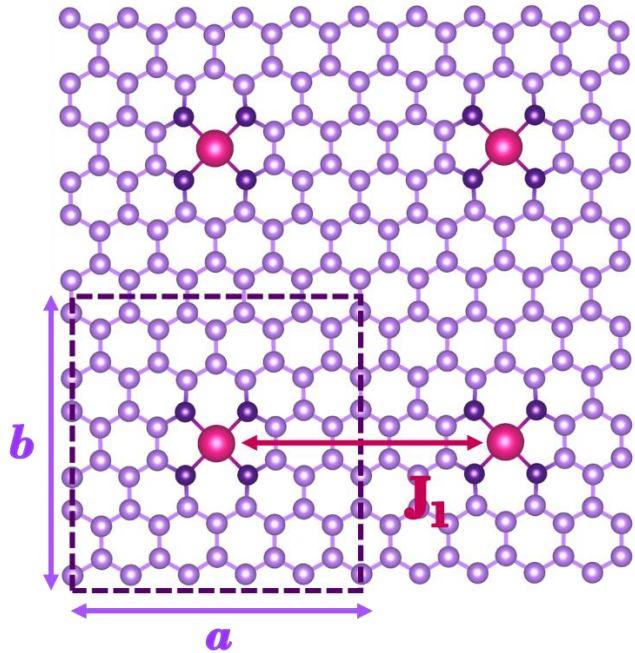

**FIG. 1.** Geometry of MN$_4$-G (M = Mn, Fe, Co, and Cu) layer. The gray, dark blue, and purple balls stand for C, N, and M atoms, respectively. J$_1$ is the exchange coupling between nearest-neighbor M atoms. M atoms are arranged in a rectangular lattice with lattice constants $a$ and $b$.

## II. Our approach to inducing strain in MN$_4$-G layers

In this section, we explain our approach to inducing strain in MN$_4$-G layers. In the first step, equilibrium structures of MN$_4$-G layers are found through optimization. Subsequently, the metal atoms in each equilibrium structure are substituted with other metal atoms. For instance, to apply strain in a FeN$_4$-G layer, the Mn atom from the optimized MnN$_4$-G structure is swapped with a Fe atom. This process leads to the creation of a strained FeN$_4$-G layer, where the Fe atom is positioned in non-equilibrium locations, effectively resulting in strain within the layer.

The strained layers are denoted by the M-X symbol, where X represents the metal atom in the equilibrium structure obtained through optimization which is the host system, and M is the guest metal atom. It should be emphasized that when M and X belong to the same metal atom, the M-X is the representative of the equilibrium structure which is obtained through optimization.

By comparing the guest-host structure with the optimized equilibrium structures, we define $\delta a = a - a_{eq}$ and $\delta b = b - b_{eq}$, where $a$ and $b$ are lattice constants in the strained structure (the guest-host structure), and $a_{eq}$ and $b_{eq}$ are lattice constants in unstrained structure (the host structure), shown in Fig. 1. These parameters measure the deviation of the lattice constants $a$ and $b$ from their equilibrium values. Accordingly, the induced strain $S_a$ and $S_b$ are defined as

$$S_{a(b)} = \frac{\delta a \ (\delta b)}{a_{eq}(b_{eq})} \times 100, \tag{1}$$

where the fraction was scaled by a factor of 100 to express the result as a percentage. A positive (negative) $S_a$ or $S_b$ means that the guest-host M-X layers are under tension (compression).

In Table 1, we have reported our results for different configurations. Based on these results, we can classify M-X layers into three distinct groups:

1) The first group includes Mn-Fe, Mn-Co, Fe-Co, Cu-Fe, and Cu-Co layers, where the induced strain causes compression of the lattice.

2) The second group comprises Fe-Mn, Fe-Cu, and all Co-X layers, where the induced strain leads to stretching of the lattice.

3) In the third group, which includes Mn-Cu and Cu-Mn layers, one of the lattice constants ($a$ or $b$) is stretched while the other is compressed.

To systematically compare the outcomes, an appropriate criterion is the change in the unit-cell area of the layers. We define $\eta = \frac{\delta A}{A_{eq}} \times 100$, with $\delta A = A - A_{eq} = ab - a_{eq}b_{eq}$. The fraction was multiplied by a factor of 100 to convert the result to a percentage value. A negative $\eta$ signifies a decrease in unit-cell area, while a positive $\eta$ indicates an increase in unit-cell area due to induced strain. We have summarized the values of $S_a$, $S_b$, and $\eta$ for various MN$_4$-G layers in Table 1.

**Table 1**: Evaluation of induced strain in various MN$_4$-G layers. The parameter $\eta$ represents the change in the unit-cell area of the layers. $S_a$ ($S_b$) is induced strain in the $a$ ($b$) lattice constant. A negative $\eta$ ($\eta < 0$), indicates lattice compression due to strain, while a positive $\eta$ ($\eta > 0$) signifies lattice stretching. $\eta = 0.00$ reflects unstrained MN$_4$-G layers.

| Layers | $S_a$ | $S_b$ | $\eta$ |
|---|---|---|---|
| Mn-Mn | ------- | ------- | 0.00 |
| Mn-Fe | -0.09635 | -0.10686 | -0.20 |
| Mn-Co | -0.19578 | -0.16312 | -0.36 |
| Mn-Cu | -0.08762 | 0.11382 | 0.03 |
| | | | |
| Fe-Fe | ------- | ------- | 0.00 |
| Fe-Mn | 0.09644 | 0.10698 | 0.20 |
| Fe-Co | -0.09953 | -0.05631 | -0.16 |
| Fe-Cu | 0.00874 | 0.22092 | 0.23 |
| | | | |
| Co-Co | ------- | ------- | 0.00 |
| Co-Mn | 0.19617 | 0.16338 | 0.36 |
| Co-Fe | 0.09963 | 0.05635 | 0.16 |
| Co-Cu | 0.10838 | 0.27739 | 0.39 |
| | | | |
| Cu-Cu | ------- | ------- | 0.00 |
| Cu-Mn | 0.08769 | -0.11369 | -0.03 |
| Cu-Fe | -0.00874 | -0.22043 | -0.23 |
| Cu-Co | -0.10826 | -0.27662 | -0.38 |

Henceforth, we will label the M-M layers and M-X layers with $\eta < 0$ and $\eta > 0$ as unstrained MN$_4$-G layers, compressed MN$_4$-G layers, and stretched MN$_4$-G layers, respectively.

**III. Effects of the strain on the electronic properties of MN$_4$-G layers**

In this section, we explore how induced strain affects the electronic properties of MN$_4$-G layers. Initially, we determine the optimized structure of unstrained layers. Similar to optimized graphene layers, where transition metal-doped vacancies create Jahn-Teller distortions around the vacancies [54-56], all MN$_4$-G unit-cells are optimized to distorted structures due to the presence of the MN$_4$ moiety. Typically, Jahn-Teller distortions partially lift $d$ orbital degeneracy in magnetic systems. However, in MN$_4$-G layers, Jahn-Teller distortion disrupts lattice symmetry and completely lifts $d$ orbital degeneracy.

Figure 2 depicts $d$ orbital energy levels of various strained ($\eta \neq 0$) and unstrained ($\eta = 0$) MN$_4$-G layers. Strain shifts $d$ orbital energy levels and changes their order. For example, in the compressed MnN$_4$-G layer, the energy levels exhibit a similar trend to those of the unstrained layer, with the only reversal occurring in the $d_{xy}$ and $d_{xz}$ (minority-spin) states.

In contrast, in the FeN$_4$-G layers, both in the compressed and in the stretched, as well as in the stretched CoN$_4$-G layer, strain significantly alters $d$ orbital energy levels order. In the stretched MnN$_4$-G layer, as well as in both the compressed and stretched CuN$_4$-G layers, the energy levels of the $d$ orbitals remain unchanged.

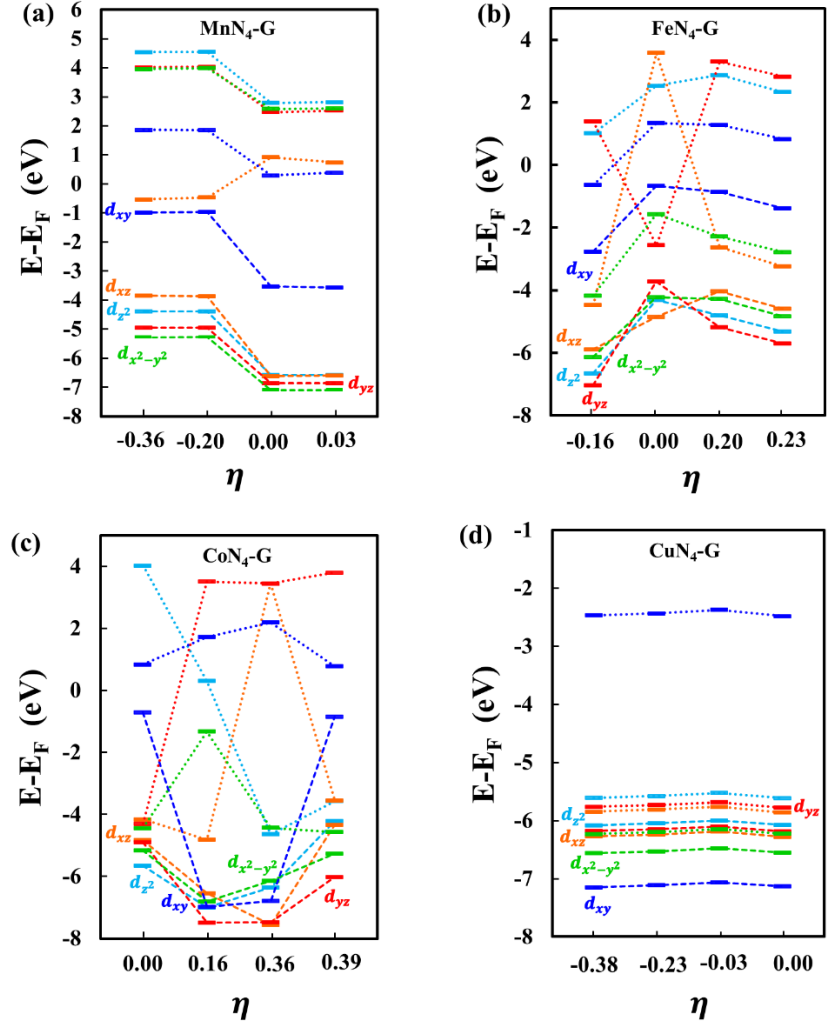

**FIG. 2.** The *d* orbital splitting in the strianed ($\eta \neq 0$) and unstrained ($\eta = 0$) MN$_4$-G layers. Panels (a) through (d) correspond to MnN$_4$-G, FeN$_4$-G, CoN$_4$-G, and CuN$_4$-G layers, respectively. The dashed (dotted) lines serve as visual guides to illustrate the energy changes of majority-spin (minority-spin) energy levels. The Fermi level is set at zero.

We have also shown the band structures of various MN$_4$-G layers in Fig. 3. As observed, the majority-spin and minority-spin bands in unstrained layers cross the Fermi level, confirming the metallic nature of the MN$_4$-G layers. This characteristic persists even under induced strain.

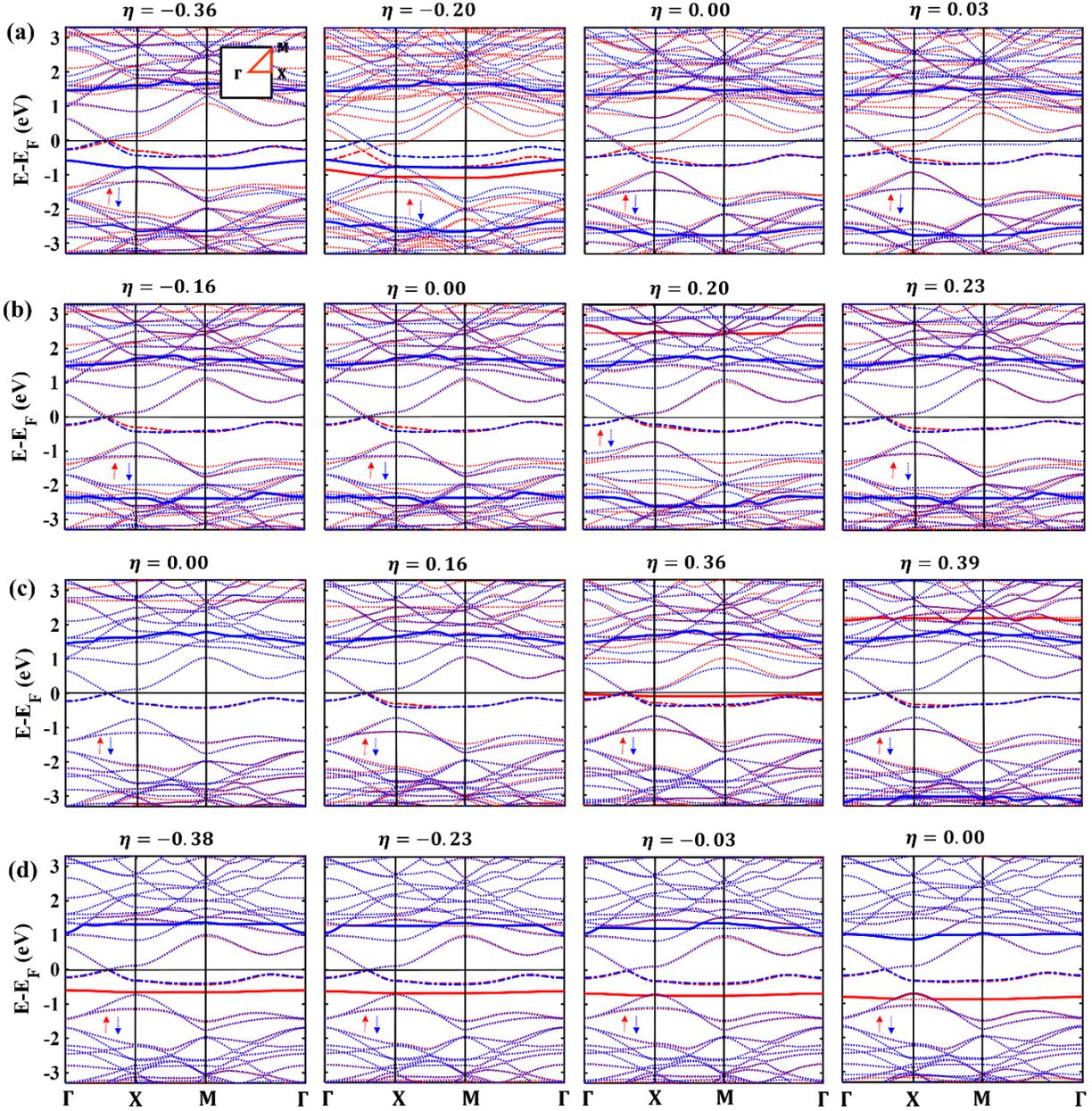

**FIG. 3.** Spin-polarized band structure of various strained and unstrained MN$_4$-G layers, for different values of $\eta$. Panels (a) through (d) correspond to MnN$_4$-G, FeN$_4$-G, CoN$_4$-G, and CuN$_4$-G layers, respectively. The red and blue thick solid lines are nearly flat bands. The red and blue thick dashed lines indicate the spin-splitting in the first valence band nearest to the Fermi level. The Fermi level is set at zero. The inset in the first frame of panel (a) illustrates the irreducible Brillouin zone of the MN$_4$-G layers.

Another interesting feature seen from Fig. 3, is the emergence of nearly flat bands. The density of states (DOS) quantifies the number of available states within a narrow energy range. Conversely, the bandwidth is associated with the orbital overlap between adjacent atoms. Consequently, flat bands exhibit a higher density of states within a small energy range compared to highly dispersive bands. The flat bands correspond to the localized $d$ orbitals of the M atom, as reported by Orellana [57] and Wu et al. [58].

When MN$_4$-G layers are subjected to strain, the number and position of flat bands are significantly impacted. For example, the number of flat bands increases in the compressed MnN$_4$-G layer with $\eta = -0.20$ and in the stretched FeN$_4$-G layer with $\eta = 0.20$. For the stretched CoN$_4$-G layers with $\eta = 0.36$ and $0.39$, a similar trend is also observed. Notably, a flat band is present at the Fermi level specifically for CoN$_4$-G layer at $\eta = 0.36$.

An additional feature observed in the band structure of MN$_4$-G layers is the spin-splitting in a band between majority-spin and minority-spin states ($\Delta_n = E_n^\uparrow - E_n^\downarrow$). The band-splitting ($\Delta_n$) arises from the exchange coupling between magnetic atoms, M. The energy difference between majority-spin and minority-spin states makes the electrons at the Fermi level to occupy different states depending on their spin direction, resulting in distinct conduction properties [59]. We have depicted in Fig. 4 the band-splitting in the first valence band nearest to the Fermi level ($\Delta_1$). Focusing initially on the unstrained MN$_4$-G layers, our results show a momentum-dependent (anisotropic) spin-splitting in the MnN$_4$-G and FeN$_4$-G layers (Fig. 4 (a, b)). Conversely, the CoN$_4$-G and CuN$_4$-G layers exhibit an almost momentum-independent (isotropic) spin-splitting, as shown in Fig. 4 (c, d).

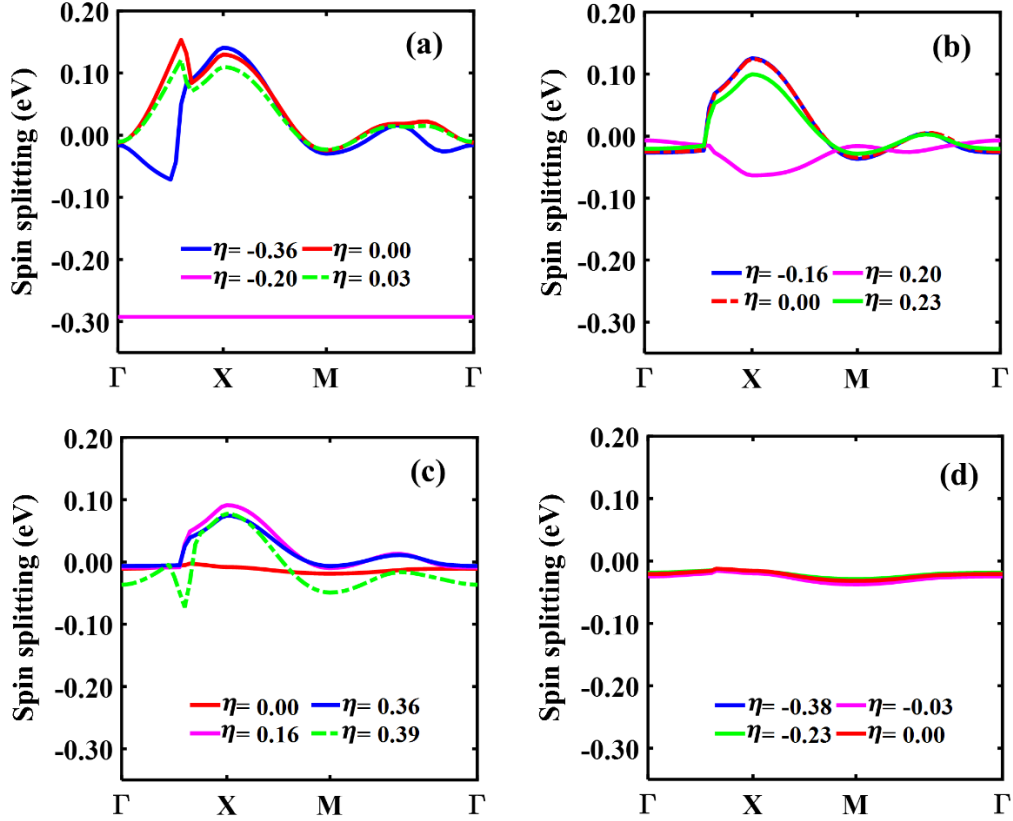

**FIG. 4.** The spin-splitting of the first valence band nearest to the Fermi level ($\Delta_1$). Panels (a) through (d) correspond to MnN$_4$-G, FeN$_4$-G, CoN$_4$-G, and CuN$_4$-G layers, respectively. Negative value of $\Delta_1$ reflects higher energy of the majority-spin states.

When MN$_4$-G layers are subjected to strain, significant changes are observed in spin-splitting. For example, in compressed MnN$_4$-G layers with $\eta = -0.20$ and in all stretched CoN$_4$-G layers, the bands exhibit isotropic and anisotropic spin-splitting, respectively, in contrast to their unstrained counterparts. The anisotropic spin-splitting in energy bands facilitates spintronic applications [60, 61]. Additionally, the magnitude of the spin-splitting increases in these cases (Fig. 4 (a, c)). Notably, in all strained CuN$_4$-G layers, no significant changes are observed in the bands exhibiting spin-splitting under applied strain.

We have also illustrated in Fig. 5 the charge density distribution near the Fermi level, $n_i(r) = n_i^\uparrow(r) + n_i^\downarrow(r)$, where $n_i^\uparrow(r)$ and $n_i^\downarrow(r)$ are respectively the spin-up and spin-down electron densities at the position $r$ in the $i$th unit-cell. In the MnN$_4$-G and CoN$_4$-G layers, the strain significantly modifies the charge density around each atom within a unit-cell. Conversely, in other layers, the local charge density exhibits no substantial variations.

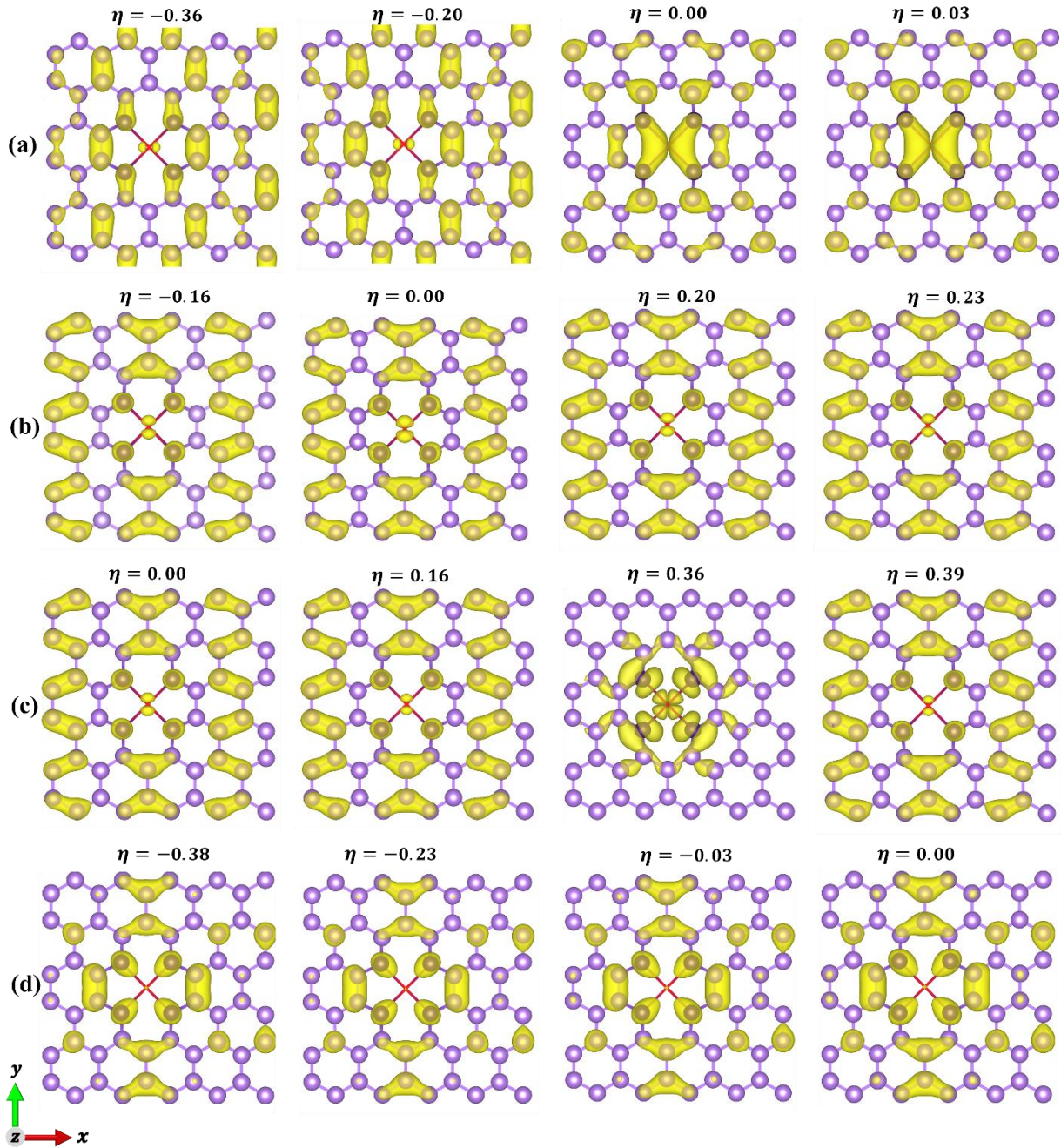

**FIG. 5.** Charge density distribution (top view: the $xy$ plane) near the Fermi level for various strained and unstrained MN$_4$-G layers. Panels (a) through (d) correspond to MnN$_4$-G, FeN$_4$-G, CoN$_4$-G, and CuN$_4$-G layers, respectively. The charge density isosurface is set at 0.0007 $e$/Å$^3$. M atoms are not shown here.

## IV. Magnetic properties of strained MN$_4$-G layers

### A. Magnetic moment

The magnetic moment of the $i$th unit-cell of a MN$_4$-G layer is given by the formula [62, 63]:

$$\mu_i = \int [n_i^\uparrow(\boldsymbol{r}) - n_i^\downarrow(\boldsymbol{r})]\, d^2r, \qquad (2)$$

where $n_i^\uparrow(\boldsymbol{r})$ and $n_i^\downarrow(\boldsymbol{r})$ are respectively the spin-up and spin-down electron densities in the $i$th unit-cell. The integral in Eq. (2) is performed locally around the M atom over the area of the $i$th unit-cell.

We have illustrated in Fig. 6 the spin density distribution ($\Delta n_i(\boldsymbol{r}) = n_i^\uparrow(\boldsymbol{r}) - n_i^\downarrow(\boldsymbol{r})$) of various strained and unstrained MN$_4$-G layers. As clearly observed, in most layers, except for the CuN$_4$-G, the spin density is highly localized on the transition metal atoms and minimal on the nitrogen atoms.

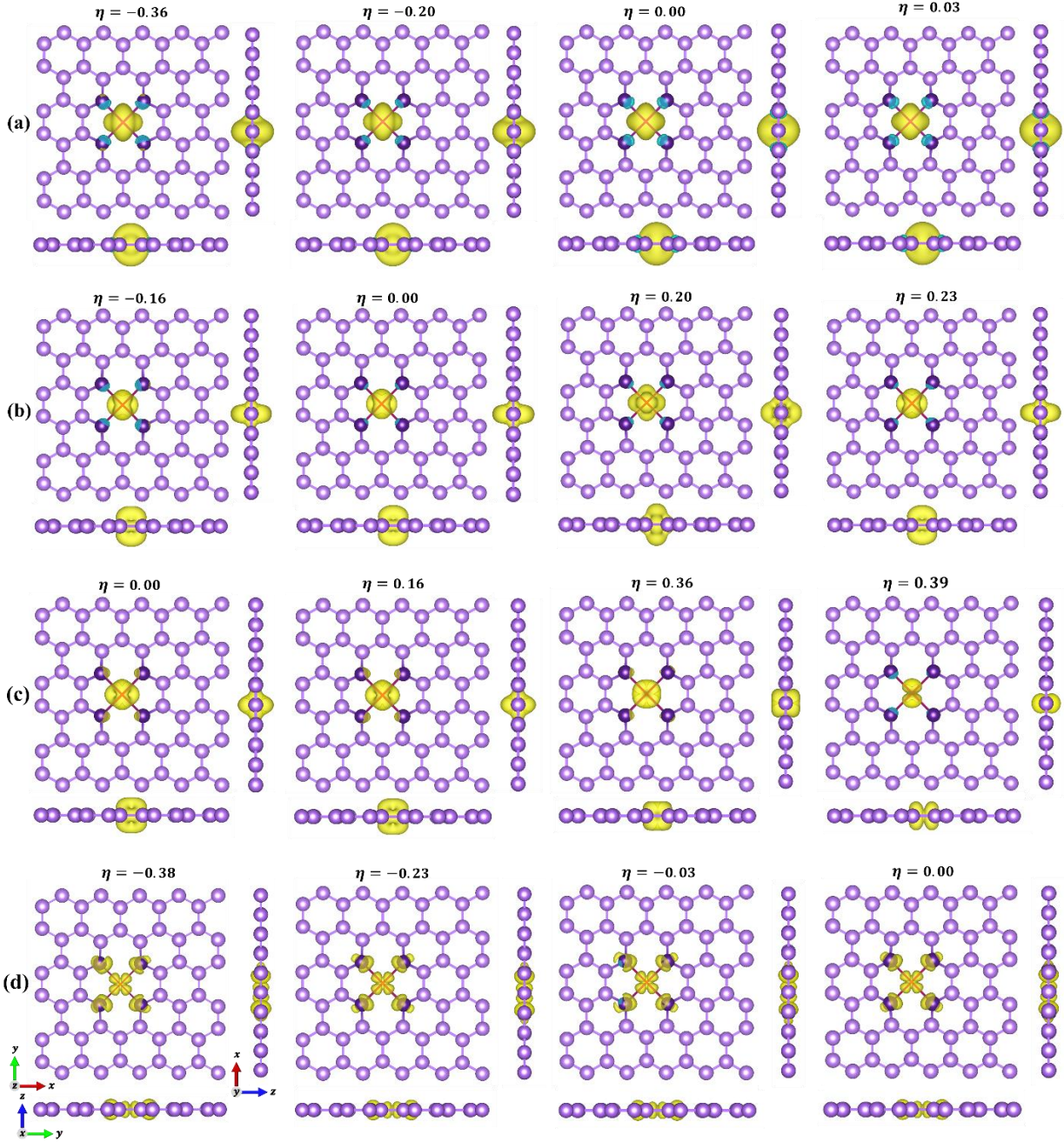

**FIG. 6.** Spin density distribution ($\Delta n(\mathbf{r})$) of various strained and unstrained MN$_4$-G layers, for different values of $\eta$. For each state three distinct projections are illustrated: the top view ($xy$-plane), the side view ($xz$-plane), and the front view ($yz$-plane). Panels (a) through (d) correspond to MnN$_4$-G, FeN$_4$-G, CoN$_4$-G, and CuN$_4$-G layers, respectively. The spin density isosurface is set at $0.008\ e/\text{Å}^3$. Yellow and cyan colors show the majority-spin density ($\Delta n > 0$) and minority-spin density ($\Delta n < 0$) distributions. M atoms are not shown here.

The shape of the spin density distribution in MN4-G layers is influenced by the contributions of various $d$ orbitals to the spin density. To explore the contribution of different $d$ orbitals, we have plotted in Fig. 7, the projected density of states (PDOS) for various layers.

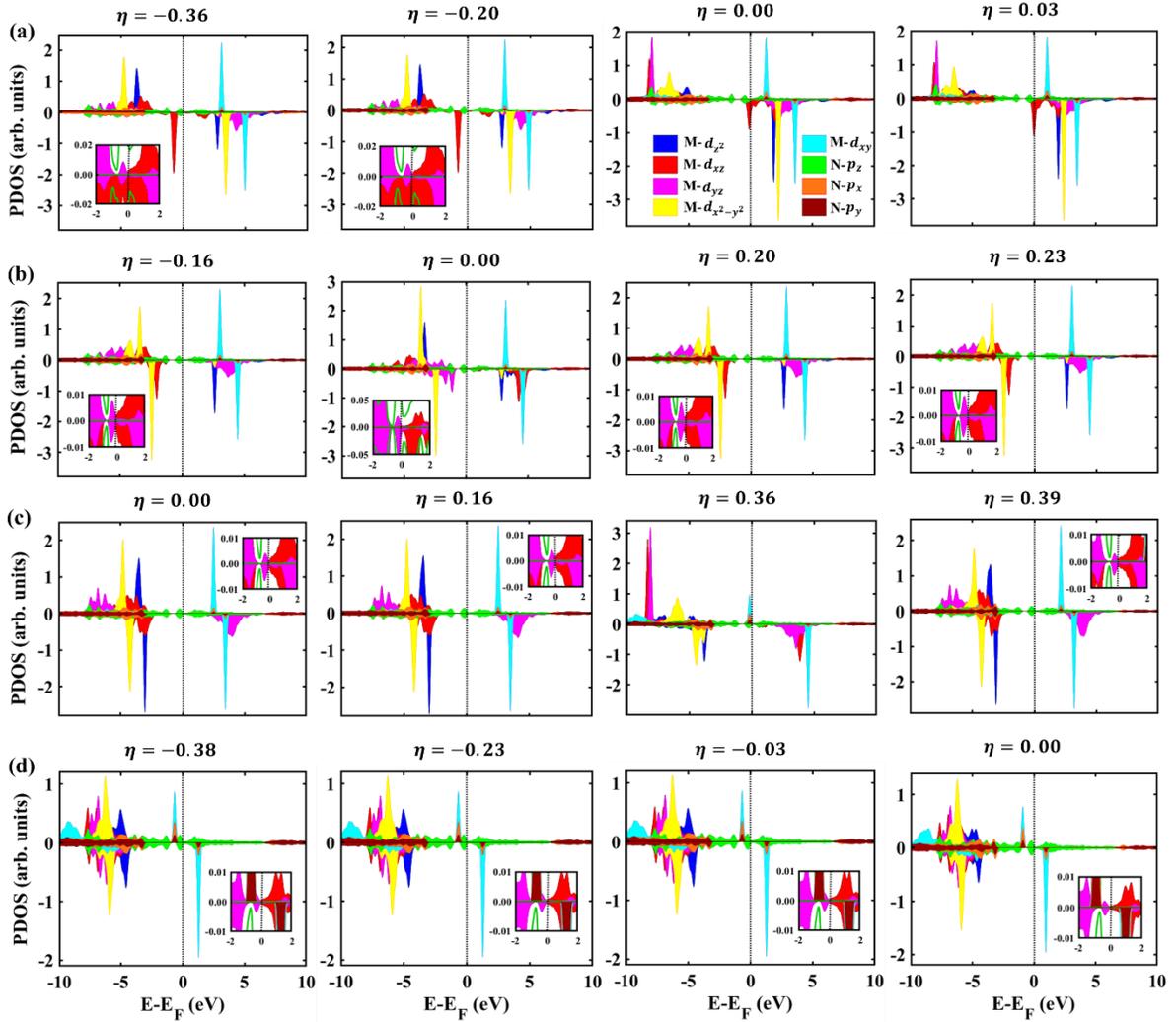

**FIG. 7.** Spin-polarized projected density of states (PDOS) of various strained and unstrained MN4-G layers. Panels (a) through (d) correspond to MnN4-G, FeN4-G, CoN4-G, and CuN4-G layers, respectively. The insets provide a magnified view of the PDOS at low energy levels. To clarify the contribution of the $d$ orbitals near the Fermi level, the inset figures show the DOS of $p_z$ orbital of N atoms as an unfilled region. The Fermi level is set at zero.

As discussed, the magnetic moment of each unit-cell predominantly originates from the $d$ orbitals of the M atom. In this respect, the magnetic moment can be expressed in terms of the PDOS on the M atoms, as $\mu = N_d^\uparrow - N_d^\downarrow$, where $N_d^\sigma = \int g_d^\sigma(E)\, dE$. Here, $g_d^\sigma(E)$ denotes spin-$\sigma =\uparrow,\downarrow$ electrons density of states of the $d$ orbitals (see Fig. 7).

In Table 2 we have reported the magnetic moment per unit-cell of various strained and unstrained MN$_4$-G layers.

**Table 2:** The magnetic moment per unit-cell of various strained and unstrained MN$_4$-G layers, for different values of $\eta$.

| MnN$_4$-G | | FeN$_4$-G | | CoN$_4$-G | | CuN$_4$-G | |
|---|---|---|---|---|---|---|---|
| $\eta$ | $\mu_i(\mu_B)$ | $\eta$ | $\mu_i(\mu_B)$ | $\eta$ | $\mu_i(\mu_B)$ | $\eta$ | $\mu_i(\mu_B)$ |
| -0.36 | 3 | -0.16 | 2 | 0.00 | 1 | -0.38 | 1 |
| -0.20 | 3 | 0.00 | 2 | 0.16 | 3 | -0.23 | 1 |
| 0.00 | 3.5 | 0.20 | 2 | 0.36 | 3 | -0.03 | 1 |
| 0.03 | 3.5 | 0.23 | 2 | 0.39 | 1 | 0.00 | 1 |

## B. Exchange coupling

One of the most common and successful microscopic models for describing the coupling between magnetic atoms in systems with inversion symmetry is the Heisenberg model [64]:

$$H = \sum_{i\neq j} J_{ij} S_i \cdot S_j , \qquad (3)$$

where $J_{ij}$ is the exchange coupling of the spins residing at sites $i$ ($S_i$) and $j$ ($S_j$). The relationship between the magnetic moment ($\mu$) and the spin (S) of the M atoms is expressed as $\mu = g\mu_B(S(S+1))^{1/2}$ where $g$ is a $g$-factor. There are several methods to estimate $J_{ij}$ within SP-DFT. The most straightforward and commonly used approach involves calculating the total energies of the $P+1$ magnetic configurations, where $P$ represents the number of distinct exchange couplings [65-67]. However, this method has several notable drawbacks: (a) it requires generating numerous distinct magnetic configurations for complex systems; (b) all configurations must maintain the same magnetic moments, which is crucial for materials near the itinerant regime; and (c) the exchange coupling obtained is a single number, making it challenging to identify which orbitals significantly influence the exchange coupling behavior and to determine the responsible mechanisms, such as direct or indirect exchange [68]. To overcome these limitations, the Green's function method is employed [69-71]. In this approach, analytical expressions have been derived for the changes in total energy due to small spin rotations via the magnetic force theorem [69, 72, 73], utilizing SP-DFT and the Heisenberg model. Using this approach, we can derive both the total exchange couplings and the contributions of individual orbitals to the overall exchange coupling from a single magnetic configuration. The approach developed by Dm. M. Korotin et al. [68] has been adapted for plane-wave methods through the formalism of Wannier functions (WFs) [71,74]. We utilized this methodology, where the transformation of plane-wave eigenstates to WFs was performed using the *wannier_ham.x* utility available with the Quantum Espresso code.

All the above-mentioned $MN_4$-G layers exhibit inversion symmetry in both strained and unstrained conditions. Moreover, the M (C and N) atoms are elements with weak (very weak) spin-orbit coupling. In this respect, the M-M exchange couplings can be incorporated in a generalized Heisenberg Hamiltonian:

$$H = \sum_{i \neq j} J_{ij} \mathbf{e}_i \cdot \mathbf{e}_j \qquad (4)$$

where $\mathbf{e}_i$ and $\mathbf{e}_j$ refers to the unit spin vectors of atoms residing at sites $i$ and $j$. A negative (positive) sign of $J_{ij}$ indicates a ferromagnetic (antiferromagnetic) coupling. In order to compute the exchange coupling $J_{ij}$, all orbitals contribution should be taken into account [75]. The couplings between different $d$ orbitals of two M atoms make a 5 × 5 matrix given by:

$$\mathcal{J}_{ij} = \begin{pmatrix} d_{z^2}d_{z^2} & d_{z^2}d_{xz} & d_{z^2}d_{yz} & d_{z^2}d_{x^2-y^2} & d_{z^2}d_{xy} \\ d_{xz}d_{z^2} & d_{xz}d_{xz} & d_{xz}d_{yz} & d_{xz}d_{x^2-y^2} & d_{xz}d_{xy} \\ d_{yz}d_{z^2} & d_{yz}d_{xz} & d_{yz}d_{yz} & d_{yz}d_{x^2-y^2} & d_{yz}d_{xy} \\ d_{x^2-y^2}d_{z^2} & d_{x^2-y^2}d_{xz} & d_{x^2-y^2}d_{yz} & d_{x^2-y^2}d_{x^2-y^2} & d_{x^2-y^2}d_{xy} \\ d_{xy}d_{z^2} & d_{xy}d_{xz} & d_{xy}d_{yz} & d_{xy}d_{x^2-y^2} & d_{xy}d_{xy} \end{pmatrix}, \tag{5}$$

where the conventional value of $J_{ij}$ corresponds to the sum of all the above matrix elements:

$$J_{ij} = \sum_{mm'} \mathcal{J}_{ij}^{mm'}, \tag{6}$$

with $m$ and $m'$ being the five $d$ orbitals. For computing the M-M exchange coupling, we employed Green's function formalism [69, 70, 71]. Within this formalism the matrix elements of the exchange coupling $\mathcal{J}_{ij}$ in Eq. (5) is expressed as:

$$\mathcal{J}_{ij}^{mm'} = -\frac{1}{2\pi}\text{Im}\int_{-\infty}^{E_F} d\varepsilon \sum_{nn'n''} \left( \Delta_i^{mn} G_{ij,\downarrow}^{nn'}(\varepsilon) \, \Delta_j^{n'n''} G_{ji,\uparrow}^{n''m'}(\varepsilon) \right), \tag{7}$$

where the sums on $n, n', n''$ run over different orbitals of the M atoms residing at sites $i$ and $j$. Here, $E_F$ is the Fermi energy of the system, $G_{ji,\uparrow}^{mm'}\left(G_{ij,\downarrow}^{mm'}\right)$ is the real-space intersite Green's function for spin-up (-down) electrons residing in orbitals $m$ and $m'$, and $\Delta_i^{mm'}$ is the local exchange splitting.

Following, we focus on the exchange coupling mechanism between the magnetic atoms in MN$_4$-G layers. We evaluate the exchange coupling of various strained MN$_4$-G layers using Eq. (6). The convergence of $J_{ij}$ with respect to the number of $k$-points is assessed with a progressively finer mesh. We have used 10×10×1 $k$-points mesh for all the layers.

**B-1) Nearest-neighbor exchange coupling:** Our results for the exchange coupling between nearest-neighbor magnetic atoms ($J_1$) in strained MN$_4$-G layers are illustrated as a plot in Fig. 8. The results of unstrained layers are also included for comparison.

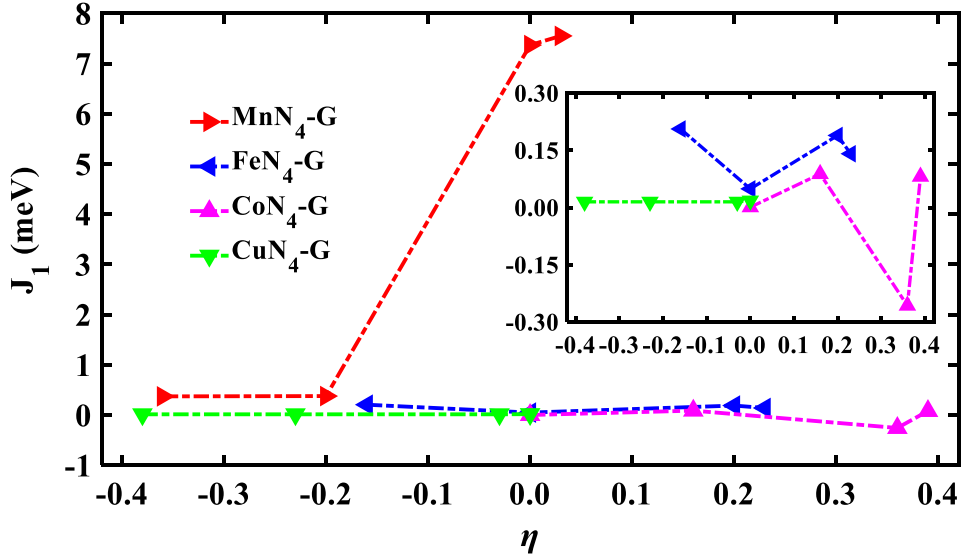

**FIG. 8.** Nearest-neighbor exchange coupling ($J_1$) in strained and unstrained $MN_4$-G layers. Except for the $CuN_4$-G layer, other layers show significant variations. Notably, in the $MnN_4$-G layer, $J_1$ decreases markedly under compression compared to other layers. The inset provides a detailed view of the $J_1$ values.

While most layers experience changes in $J_1$, the magnitude of these changes is considerably greater in the compressed $MnN_4$-G layers. The variation of $J_1$ is attributed to the lattice distortions induced by strain, leading to significant modifications in electronic properties. These modifications include changes in the energy level splitting of the $d$ orbitals and the PDOS of the M atoms, as discussed in the previous section.

To examine the contribution of different $d$ orbitals to the exchange coupling $J_1$, we computed all the matrix elements of $J_1$. Our results show that $J_1$ is diagonal with all off-diagonal elements being zero. We have illustrated the diagonal matrix elements of $J_1$, for various strained $MN_4$-G layer, in Fig. 9. For comparison, the results for the unstrained layers are also presented. As clearly observed, in the unstrained $MnN_4$-G layer, the $d_{xz}$ orbitals predominantly contribute to $J_1$. Under compression, the overlap of $d_{xz}$ orbitals significantly decreases, resulting in a sharp reduction of $J_1$ (see Fig. 9 (a)).

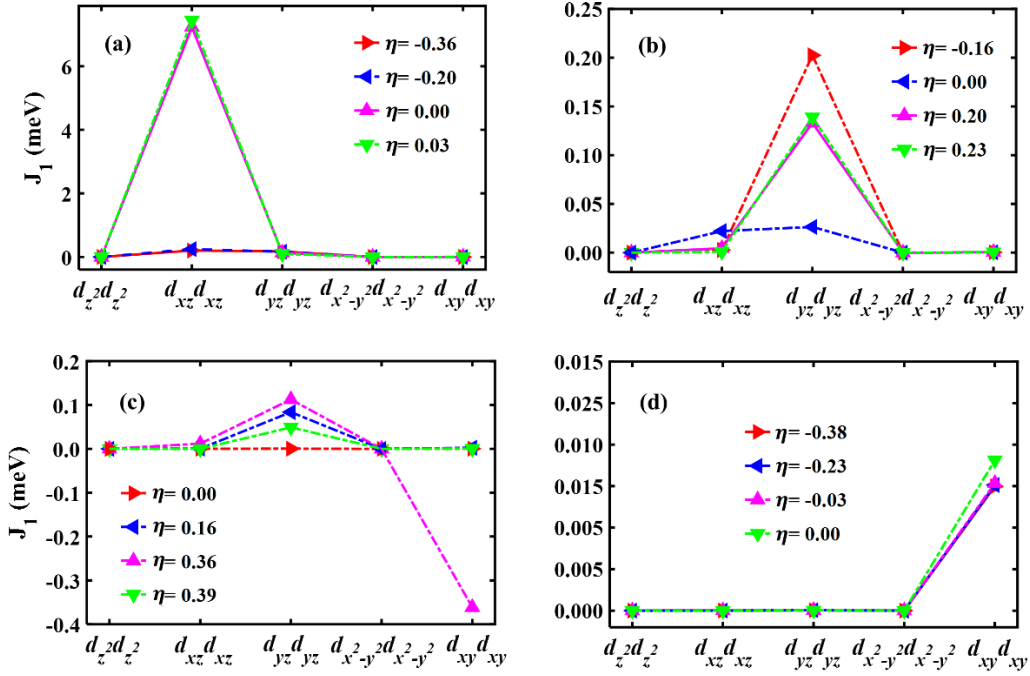

**FIG. 9.** The diagonal matrix elements of $J_1$ in various strained and unstrained MN$_4$-G layers. Panels (a) through (d) correspond to MnN$_4$-G, FeN$_4$-G, CoN$_4$-G, and CuN$_4$-G layers, respectively.

In the FeN$_4$-G and CoN$_4$-G layers, the $d_{yz}$ orbital plays the role. As clearly observed, in the strained FeN$_4$-G layers, both with $\eta > 0$ and $\eta < 0$, the overlap of the $d_{yz}$ orbital increases. In CoN$_4$-G layer, both the $d_{yz}$ and $d_{xy}$ orbitals contribute effectively (Fig. 9 (c)). While the unstrained CoN$_4$-G layer shows a weak $J_1$, this strength increases under strain. At $\eta = 0.36$ the $d_{xy}$ orbital contribution significantly rises, resulting in a much large $J_1$ compared to unstrained layer. In CuN$_4$-G layer, the $d_{xy}$ orbital predominantly contributes to the $J_1$ coupling, and its contribution remains largely unaffected by the induced strain.

**B-2) Exchange coupling mechanism:** To understand the effect of induced strain on the exchange coupling mechanism, we have examined the changes in the behavior of the exchange coupling $J_r$ with respect to the M-M separation distance $r$. We have demonstrated in our previous study [52] that in the unstrained MN4-G layers (except for the CuN4-G layer) the exchange mechanism is RKKY, and the coupling $J_r$ decreases oscillatory by increasing $r$ (see Fig. 10 (a, b, c), the unstrained ($\eta = 0$) cases). In contrast, the exchange mechanism in the unstrained CuN4-G layer is superexchange which is short-range and decays rapidly by increasing the Cu-Cu separation distance. The unstrained CuN4-G layer can be effectively served as a one-dimensional antiferromagnetic chain [52].

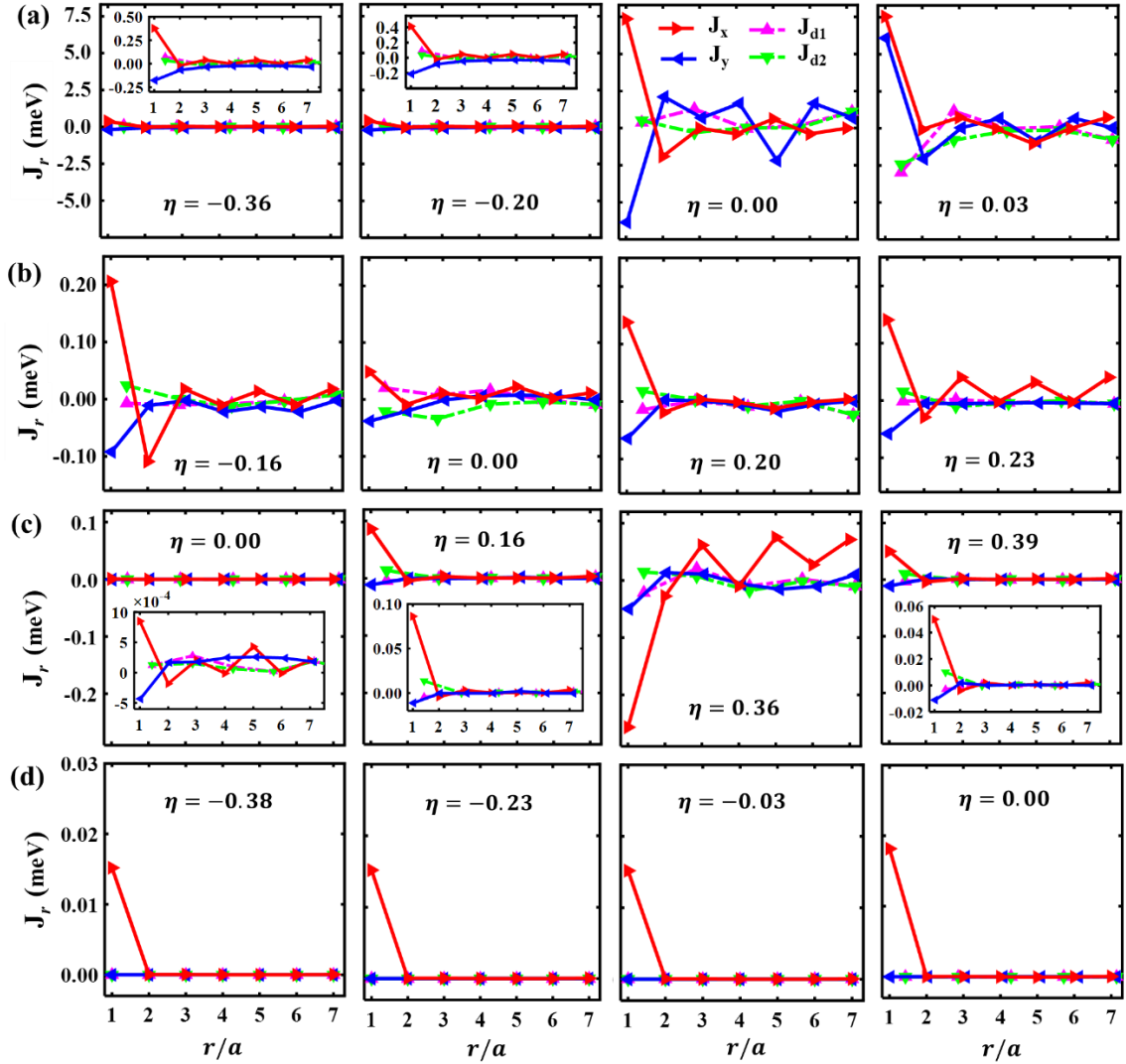

**FIG. 10.** The excahnge coupling in strained and unstrained MN4-G layers as a function of M-M separation distance ($r/a$) along $x$, $y$, d1, and d2 directions (see Fig. 12 in appendix). Panels (a) through (d) correspond to MnN4-G, FeN4-G, CoN4-G, and CuN4-G layers, respectively. The insets are zoom-in view of the coupling along the $y$-axis.

Interestingly, the exchange coupling mechanism in all MN$_4$-G (M=Mn, Fe, Co, and Cu) layers remains unaffected by strain, as illustrated in Fig. 10. Strain modifies the amplitude of RKKY oscillations in the MN$_4$-G layers (except for the CuN$_4$-G layer). The amplitude of RKKY oscillations is significantly suppressed in the MnN$_4$-G layers under compressive strain. Conversely, the amplitude of RKKY oscillations is increased in the FeN$_4$-G and CoN$_4$-G layers, with a notable enhancement in FeN$_4$-G under compressive strain and in CoN$_4$-G at $\eta = 0.36$. The variation in the amplitude of RKKY oscillations is attributed to changes in electron density. The strength of RKKY coupling is determined by the DOS of conduction electrons at the Fermi level and the overlap between the $d$ orbitals of magnetic atoms and conduction electrons.

To analyze the effects of the induced strain on the decay rate of the RKKY couplings, we fitted the data points of exchange couplings along $x$, $y$, d1, and d2 directions with the following oscillatory function [52].

$$J_r = J_0 \left[\frac{\cos(2ak_F r)}{(2ak_F r)^n}\right], \tag{8}$$

where $k_F$, $J_0$, $a$ and $n$ are the Fermi wave number, a constant with a unit of energy, the lattice constant along the $x$ direction, and the decay rate of the RKKY coupling, respectively. In Tables 3, 4 and 5 (see appendix) we have reported the fitting parameters.

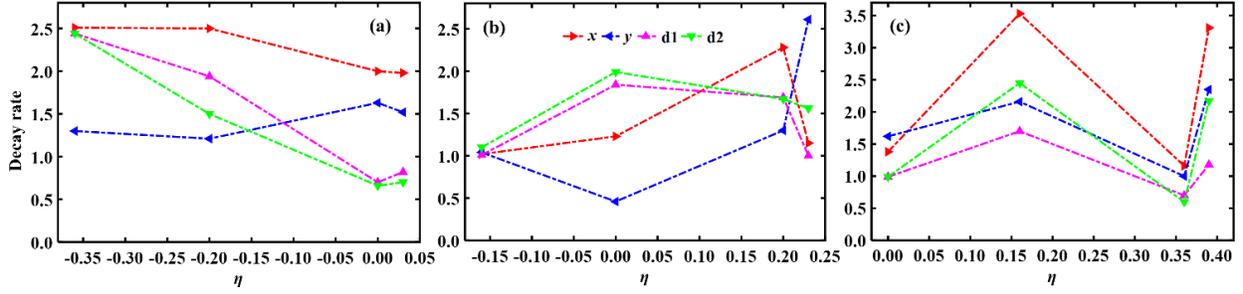

**FIG. 11.** The decay rate of the RKKY coupling ($n$) in various strained and unstrained MN$_4$-G layers. The $x$, $y$, and two main diameters (d1, d2) directions are illustrated in Fig. 12 in appendix. Panels (a) through (c) correspond to MnN$_4$-G, FeN$_4$-G, and CoN$_4$-G layers, respectively.

The decay rate of the RKKY coupling ($n$) in the MN$_4$-G layers with M=Mn, Fe, and Co, along the directions $x$, $y$, d1 and d2 are shown in Fig. 11. It is clear that the decay rate of the MnN$_4$-G layer under compressive strain ($\eta < 0$) increases along the $x$, d1, and d2 directions, while decreasing along the $y$ direction. In contrast, the decay rate in the MnN$_4$-G layer under stretched strain ($\eta > 0$) remains almost unaffected in all directions. In the compressed FeN$_4$-G layer, the decay rate increases along the $y$ direction and decreases along the $x$, d1, and d2 directions. In contrast, in stretched FeN$_4$-G layers, the decay rate consistently increases along the $y$ direction and decreases along the d1 and d2 directions. Along the $x$ direction, the decay rate initially increases at lower $\eta$ but eventually decreases to a value lower than that of the unstrained layer. The decay rate in the CoN$_4$-G layer exhibits an oscillatory behavior. It initially increases with increasing $\eta$, significantly decreases at $\eta = 0.36$, and then increases again, reaching a value at $\eta = 0.39$ that exceeds the rate in the unstrained layer.

Applying strain allows us to modulate the coupling rate. In certain layers, strain decreases the rate, while in others, it increases the rate. Modulating the coupling rate through applied stress enables the design of systems with tailored magnetic properties. A slow decay rate for the RKKY coupling suggests long-range coupling which makes these layers promising for spintronic applications.

To calculate the Fermi wave number ($k_F^{Cal}$) for various strained and unstrained MN$_4$-G layers with M= Mn, Fe, and Co, we use the following

$$k_F^{Cal} = k_F^\uparrow + k_F^\downarrow, \quad k_F^\sigma = \left(2\pi \frac{N_e^\sigma}{A}\right)^{1/2}, \tag{9}$$

where $N_e^\sigma$ refers to the number of electrons with spin $\sigma = \uparrow, \downarrow$ located near the Fermi level, and A denotes the unit-cell area. The magnitude of the Fermi wave number obtained by fitting ($k_F^{Fit}$) is almost in agreement with calculated Fermi wave number, as detailed in Table 6 in appendix.

## V. Computational details

To calculate the spin-polarized (SP) electronic structure we used the density functional theory (DFT) framework, as implemented in the Quantum Espresso code [76]. We utilize the Perdew-Burke-Ernzerhof revised for solids (PBEsol) [77] exchange-correlation functional within the generalized gradient approximation [78]. The ion-electron coupling is treated by the projector augmented wave method [79]. For sampling of the Brillouin zone, Γ-centered Monkhorst-Pack [80] $k$-point meshes of $2 \times 2 \times 1$ are used for all MN$_4$-G layers, and an energy cutoff of 1088.46 eV is adopted for the plane-wave basis.

The structure optimization stopped when the residual force on each atom was <0.026 eV/Å and energy was <0.014 eV. We have used the Hubbard parameter for the MN$_4$-G layers from Ref. [52].

## VI. Summary and concluding remarks

In summary, we have conducted a comprehensive investigation into the effects of strain on the exchange coupling and magnetic properties of MN$_4$-G (M = Mn, Fe, Co, Cu) layers using ab-initio DFT and the Green's function method. To induce strain, we employed a novel approach by embedding atoms from one layer within the equilibrium structure of another, generating either tensile or compressive strain.

Our results underscore the pivotal role of strain in modulating the electronic structure of $MN_4$-G layers. Strain induces shifts in the d-orbital energy levels and alters their order, leading to significant modifications in the band structure. These include spin-splitting and the emergence of nearly flat energy bands, suggesting a complex interplay between mechanical deformation and electronic properties. Flat bands near the Fermi level are of particular interest due to their potential to support exotic quantum states, such as superconductivity. Strain-dependent variations are especially pronounced in $CoN_4$-G layers, which exhibit a transition from isotropic to anisotropic spin-splitting configurations. In contrast, $CuN_4$-G layers remain largely unaffected by strain, preserving their unique electronic characteristics.

The strain-induced changes in electronic structure are strongly correlated with magnetic behavior. While the RKKY mechanism predominantly governs exchange coupling in most $MN_4$-G layers, $CuN_4$-G layers exhibit a distinct superexchange mechanism. These layers behave as decoupled one-dimensional spin chains with short-range Heisenberg coupling. Notably, we demonstrated that a tensile strain of +0.36% in the CoN4-G layer significantly enhances the strength and amplitude of RKKY oscillations while reducing their decay rate, enhancing its potential for spintronic applications. Conversely, the $CuN_4$-G layer's magnetic properties remain invariant under strain, preserving its decoupled spin chain behavior.

Overall, our study provides critical insights into the strain-dependent tunability of magnetic interactions in $MN_4$-G layers. These findings highlight the potential of strain engineering as a powerful tool for tailoring the electronic and magnetic properties of 2D materials, paving the way for the design of next-generation spintronic devices. Future studies could extend these results to explore practical implementations and refine control mechanisms for optimized performance in spintronics.

**ACKNOWLEDGMENTS**

F.N. and J. A. are grateful to the *Institute for Advanced Studies in Basic Sciences* for financial support through research Grant No. G2023IASBS32604 and No. GIASBS12969, respectively.

# Appendix

This part includes Fig. 12 showing the directions for calculating $J_r$ and Tables 3, 4, 5, and 6 with the fitting parameters for $J_r$.

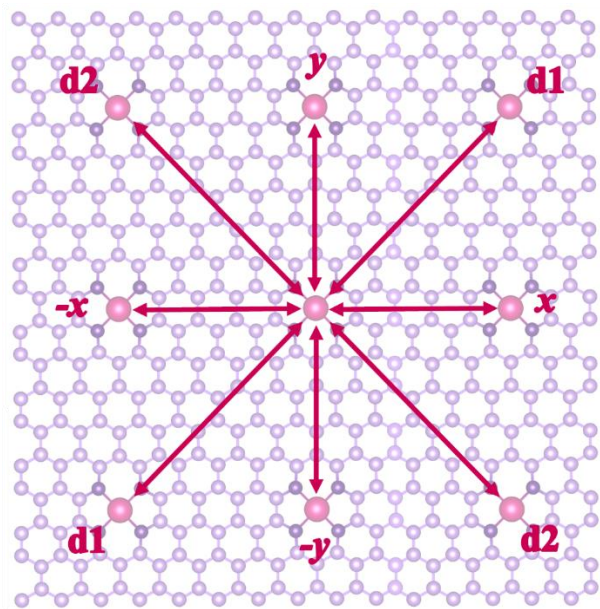

**FIG. 12.** The exchange couplings of the M atom located at the origin of a unit-cell with the neighboring atoms are shown with the purple double-head arrows along $\pm x$, $\pm y$, and the two main diameter (d1, d2) directions.

**Table 3:** Fitting parameters of $J_r$ in Eq. (8), for MnN$_4$-G layer under the strain. $k_F = [(k_F^x)^2 + (k_F^y)^2]^{1/2}$ is the Fermi wave number, $k_F^x$ and $k_F^y$ are components of the Fermi wave vector; and $J_0$ is a constant along $\pm x$, $\pm y$, and the two main diameter (d1, d2) directions.

| MnN$_4$-G | Direction | $J_0$ (meV) | $k_F^x$ (Å$^{-1}$) | $k_F^y$ (Å$^{-1}$) | $k_F$ (Å$^{-1}$) |
|---|---|---|---|---|---|
| $\eta = -0.36$ | x | -5.548 | 0.118 | ------ | ------ |
|  | y | -0.003 | ------ | 0.002 | ------ |
|  | d1 | -3.939 | ------ | ------ | 0.118 |
|  | d2 | -3.949 | ------ | ------ | 0.118 |
| $\eta = -0.20$ | x | -6.313 | 0.121 | ------ | ------ |
|  | y | -0.007 | ------ | 0.002 | ------ |
|  | d1 | -2.210 | ------ | ------ | 0.121 |
|  | d2 | -0.900 | ------ | ------ | 0.121 |
| $\eta = 0.00$ | x | -75.610 | 0.123 | ------ | ------ |
|  | y | 41.040 | ------ | 0.125 | ------ |
|  | d1 | -4.402 | ------ | ------ | 0.175 |
|  | d2 | -1.570 | ------ | ------ | 0.175 |
| $\eta = 0.03$ | x | -118.600 | 0.150 | ------ | ------ |
|  | y | -53.000 | ------ | 0.145 | ------ |
|  | d1 | -15.560 | ------ | ------ | 0.209 |
|  | d2 | -8.486 | ------ | ------ | 0.209 |

**Table 4:** Fitting parameters of $J_r$ in Eq. (8), for FeN$_4$-G layer under the strain. Other parameters are detailed in the caption of Table 3.

| FeN$_4$-G | Direction | $J_0$ (meV) | $k_F^x$ (Å$^{-1}$) | $k_F^y$ (Å$^{-1}$) | $k_F$ (Å$^{-1}$) |
|---|---|---|---|---|---|
| $\eta = -0.16$ | x | -0.600 | 0.117 | ------ | ------ |
| | y | 0.361 | ------ | 0.150 | ------ |
| | d1 | -0.100 | ------ | ------ | 0.190 |
| | d2 | 0.200 | ------ | ------ | 0.190 |
| $\eta = 0.00$ | x | -0.164 | 0.125 | ------ | ------ |
| | y | -0.029 | ------ | 0.022 | ------ |
| | d1 | -0.143 | ------ | ------ | 0.127 |
| | d2 | -1.300 | ------ | ------ | 0.127 |
| $\eta = 0.20$ | x | -1.46 | 0.108 | ------ | ------ |
| | y | -0.084 | ------ | 0.037 | ------ |
| | d1 | 0.188 | ------ | ------ | 0.114 |
| | d2 | -0.310 | ------ | ------ | 0.114 |
| $\eta = 0.23$ | x | -0.450 | 0.125 | ------ | ------ |
| | y | 0.753 | ------ | 0.091 | ------ |
| | d1 | 0.191 | ------ | ------ | 0.155 |
| | d2 | 0.319 | ------ | ------ | 0.155 |

**Table 5:** Fitting parameters of $J_r$ in Eq. (8), for CoN$_4$-G layer under the strain. Other parameters are detailed in the caption of Table 3.

| CoN$_4$-G | Direction | $J_0$(meV) | $k_F^x$(Å$^{-1}$) | $k_F^y$(Å$^{-1}$) | $k_F$(Å$^{-1}$) |
|---|---|---|---|---|---|
| $\eta = 0.00$ | x | -0.004 | 0.129 | ------ | ------ |
| | y | -0.0004 | ------ | 0.029 | ------ |
| | d1 | -0.001 | ------ | ------ | 0.132 |
| | d2 | -0.001 | ------ | ------ | 0.132 |
| $\eta = 0.16$ | x | 48.460 | 0.223 | ------ | ------ |
| | y | 0.099 | ------ | 0.097 | ------ |
| | d1 | 0.240 | ------ | ------ | 0.243 |
| | d2 | -3.721 | ------ | ------ | 0.243 |
| $\eta = 0.36$ | x | 0.833 | 0.099 | ------ | ------ |
| | y | -0.100 | ------ | 0.046 | ------ |
| | d1 | 0.109 | ------ | ------ | 0.109 |
| | d2 | -0.109 | ------ | ------ | 0.109 |
| $\eta = 0.39$ | x | 19.65 | 0.221 | ------ | ------ |
| | y | 0.131 | ------ | 0.111 | ------ |
| | d1 | 0.047 | ------ | ------ | 0.247 |
| | d2 | -1.391 | ------ | ------ | 0.247 |

**Table 6:** Calculated ($k_F^{Cal}$) and fitted ($k_F^{Fit}$) Fermi wave numbers for various strained and unstrained MN$_4$-G layers under the strain.

| MnN$_4$-G | | | FeN$_4$-G | | | CoN$_4$-G | | |
|---|---|---|---|---|---|---|---|---|
| $\eta$ | $k_F^{Cal}$(Å$^{-1}$) | $k_F^{Fit}$(Å$^{-1}$) | $\eta$ | $k_F^{Cal}$(Å$^{-1}$) | $k_F^{Fit}$(Å$^{-1}$) | $\eta$ | $k_F^{Cal}$(Å$^{-1}$) | $k_F^{Fit}$(Å$^{-1}$) |
| -0.36 | 0.112 | 0.118 | -0.16 | 0.106 | 0.190 | 0.00 | 0.105 | 0.132 |
| -0.20 | 0.100 | 0.121 | 0.00 | 0.120 | 0.127 | 0.16 | 0.149 | 0.243 |
| 0.00 | 0.125 | 0.175 | 0.20 | 0.101 | 0.114 | 0.36 | 0.139 | 0.109 |
| 0.03 | 0.137 | 0.209 | 0.23 | 0.122 | 0.155 | 0.39 | 0.132 | 0.247 |